\documentclass[a4paper,fleqn]{cas-dc}
\usepackage{caption}
\usepackage[authoryear,longnamesfirst]{natbib}
\usepackage{titlesec}
\usepackage{float}
\usepackage{tabularx}
\usepackage{titlesec}
\usepackage{caption}
\usepackage[utf8]{inputenc}
\usepackage{balance}
\usepackage{newunicodechar}
\usepackage{xr-hyper}  
\usepackage{hyperref}
\newunicodechar{−}{-}
\usepackage{titlesec}

\titleformat{\section}[block]
  {\Large\bfseries}
  {\thesection}
  {0.7em}
  {}                     

\titlespacing{\section}{0pt}{*2}{*1}

\titleformat{\subsection}[block]
  {\large\bfseries}
  {\thesubsection}
  {0.7em}
  {}

\titlespacing{\subsection}{0pt}{*1.5}{*0.8}

\titleformat{\subsubsection}[block]
  {\normalfont\itshape\normalsize}
  {\thesubsubsection}
  {0.7em}
  {}
\titlespacing{\subsubsection}{0pt}{*1.2}{*0.6}

\def\tsc#1{\csdef{#1}{\textsc{\lowercase{#1}}\xspace}}
\tsc{WGM}
\tsc{QE}
\tsc{EP}
\tsc{PMS}
\tsc{BEC}
\tsc{DE}

\begin{document}
\let\WriteBookmarks\relax
\def\floatpagepagefraction{1}
\def\textpagefraction{.001}


\shortauthors{Vivek Chowdhury et~al.}

\title [mode = title]{Strain-Tunable Spin Filtering and Valley Splitting Coexisting with Anomalous Hall Effect in 2D Half-Metallic VSe$_2$/VN Heterostructure: Toward a Unified Spintronic-Valleytronic Platform}



%
\author[1]{Vivek Chowdhury}[type=editor,
                        orcid=https://orcid.org/0009-0006-6180-8398]



\ead{vivekchybuet031@gmail.com}

\credit{Conceptualization, Methodology, Visualization, Software, Investigation, Writing – original draft}



\affiliation[1]{organization={Department of Electrical and Electronic Engineering},
    addressline={Bangladesh University of Engineering and Technology}, 
    city={Dhaka},
    postcode={1205}, 
    country={Bangladesh}}


\author[1]{Ahmed Zubair}[orcid=0000-0002-1833-2244
   ]
\cormark[1]
\ead{ahmedzubair@eee.buet.ac.bd}

\credit{Conceptualization, Methodology, Visualization, Resources, Funding acquisition, Writing – original draft, Writing – review \& editing, Supervision}





\shorttitle{VSe$_2$/VN Spintronics and Valleytronics}


\shortauthors{V. Chowdhury and A. Zubair}

\cortext[cor1]{Corresponding author}

\begin{abstract}
Rapid progress in valleytronics and spintronics is limited by the scarcity of two-dimensional materials that simultaneously provide robust valley splitting and strong spin selectivity. Here we showed that a van der Waals heterostructure (VSe$_2$/VN) built from hexagonal VSe$_2$ and hexagonal VN addressed this gap. Using first-principles density functional theory, phonon, \textit{ab initio} molecular dynamics stability tests, Bader charge analysis, and Wannier-based Berry-curvature calculations, we demonstrated an energetically and dynamically stable heterostructure that exhibited interlayer charge transfer and a work function intermediate between the constituent monolayers. The electronic structure showed a small indirect PBE gap (108.9~meV), with HSE06 indicating a half-metallic tendency; a sizable conduction-band valley splitting ($\mathrm{\Delta C_{KK'}}=22.9~\mathrm{meV}$ for spin-up and  $\mathrm{\Delta C_{KK'}}=61.3~\mathrm{meV}$ for spin-down); and pronounced spin asymmetry, where the spin-down channel showed a wide semiconducting gap (0.64~eV) while the spin-up channel was nearly gapless. These features yielded a high zero-strain spin-filter efficiency $\mathrm{P}=75.4\%$, tunable to $82.5\%$ under $+4\%$ biaxial tensile strain. The heterostructure also supported non-zero, valley-contrasting Berry curvature, and a large anomalous Hall conductivity (peak $\sigma_{xy}=568.33~\mathrm{S/cm}$). Importantly, mean-field estimation placed the ferromagnetic Curie temperature near room temperature at zero strain ($\mathrm{T_C}=284.04~\mathrm{K}$), while $\mathrm{T_C}$ decreased to $183.9~\mathrm{K}$ at $+4\%$ strain, the magnetic order remained robust to cryogenic temperatures, providing a beneficial tuning knob to balance spin-filter performance with thermal stability in device-relevant regimes. These results identified VSe$_2$/VN as a practical, strain-tunable platform for integrated valleytronic, spintronic devices, and for exploring anomalous Hall and valley-dependent transport phenomena.
\end{abstract}






\begin{keywords}
\sep Valleytronics \sep strain \sep  Spin-Filtering \sep  Valley Splitting \sep 2D Half-Metallic VSe$_2$/VN \sep Anomalous Hall Effect \sep Unified Spintronic-Valleytronic Platform \sep 2D TMD \sep Heterostructure
\end{keywords}

\maketitle

\section{Introduction}

The failure of Dennard scaling has stopped the amazing progress of Moore's law, where voltage no longer scales with feature size. This has pushed power densities to a power wall and made a lot of transistors stay inactive as dark silicon~\cite{dennard2003design}.  As wire delays, leakage, and heat dissipation dominate, further clock scaling yields diminishing returns and energy efficiency becomes the primary constraint for information processing~\cite{horowitz2005scaling}. Spintronics addresses these challenges by utilizing the electron's spin as an information carrier, facilitating nonvolatile, energy efficient state variables, and efficient charge–spin conversion, which are foundational to technologies ranging from magnetic tunnel junction (MTJ) based memory/logic to spin–orbit-torque devices \cite{hirohata2020review,vzutic2004spintronics}. The MTJ is essentially the tunneling analog of the giant magnetoresistance (GMR) spin-valve, but with the non-magnetic metal spacer replaced by an insulator. In MgO (001) barrier, GMR has been exhibited both theoretically and experimentally~\cite{butler2001spin,inomata2008highly}. Half-metallic ferromagnets (HMFs) and spin-filtering 2D materials are also another domain of spintronics.  Theoretically, HMFs provide 100\% spin polarization near the Fermi level. Heusler alloys like Co$_2$MnSi, Co$_2$FeSi, Manganite perovskites like La$_{0.7}$Sr$_{0.3}$MnO$_3$ and CrO$_2$ are some classes of HMFs~\cite{xie2025emerging}. Spin-filtering uses the idea of ferromagnetic insulating barrier where electrons with spins aligned to the barrier’s magnetization tunnel more readily than opposite spins. For example, ferrimagnetic CoFe$_2$O$_4$ barrier can be used as room-temperature spin-filtering device~\cite{inomata2008highly}. By using few-layer barriers and graphene contacts giant tunneling magnetoresistance (TMR) can be shown that grows with the number of CrI$_3$ layers~\cite{song2018giant}.

In contrast, valleytronics leverages the momentum-space valley index and Berry curvature in 2D crystals to generate and control dissipationless transverse responses offering an orthogonal route to encode and manipulate information \cite{schaibley2016valleytronics}.  Identification of efficient valleytronic systems is essential to supersede modern charge or spin-based semiconductor devices \cite{schaibley2016valleytronics}.  By manipulating the valley degree of freedom, energy degeneracy can be broken in two unequal valleys, considering the strong spin-orbit coupling (SOC) \cite{nebel2013electrons}. Different methods have been introduced to enhance the valley degeneracy so far, including time reversal symmetry (TRS) violation \cite{mekonnen2018dopant} using magnetic doping \cite{zhang2019valley}, applied external magnetic field \cite{zhai2019spin}, magnetic proximity effects \cite{zhong2020layer}, and circularly polarized light pump \cite{mak2019probing}. Magnetic doping creates lattice distortion or impurity scattering, and the magnetic field can introduce only a small valley splitting of 0.1-0.2 meV/T \cite{henriques2020excitonic}. Besides, circularly polarized light pumping is not suitable for valleytronic applications due to small carrier lifetime (\textasciitilde 1ns) and lack of control \cite{catarina2020magneto}. Transition-metal dichalcogenides (TMDs) are proven to be very successful for the application of valleytronic, spintronic, optoelectronic and biomedical devices \cite{REN201970,Shihab2022ACSOmega,ZHAO2020144367, Khaled2025NA}. Moreover, they have the features of a tunable bandgap within the range of 1 to 2 eV induced by strong photoluminescence and large binding energy due to the excitonic effect \cite{doi:10.1021/acsnano.1c01220}. Especially TMDs with H-phase have intrinsic broken spatial inversion symmetry (SIS), strong SOC, and large spin valley splitting at $\mathrm{K}$/$\mathrm{K'}$ valley, which make them potential candidates for valleytronics \cite{xiao2012coupled,yao2008valley, Samiul2025APSUSC}.

The 2D magnetic substrate forming a van der Waals (vdW) heterostructure with 2D layered materials can eliminate impurity scattering, which makes it a good choice for fabricating valleytronic devices \cite{bian2022tunable,liang2017magnetic}. The magnetic proximity effect enables monolayer MoTe$_2$ deposited on ferromagnetic substrate EuO a large valley splitting of 300 meV \cite{zhang2016large}. Experimental findings showed that valley splitting of monolayer WSe$_2$ can be increased with the help of magnetic substrate EuS \cite{zhao2017enhanced}. Heterostructure of WSe$_2$ and CrI$_3$ showed a considerable amount of valley splitting \cite{seyler2018valley}. Besides, interfacial orbital hybridization allows a larger valley splitting in stanene/CrI$_3$ heterostructure \cite{zhai2019spin}. Giant valley splitting (412 meV) was observed in 2D WS$_2$/h-VN heterostructure \cite{Ke}. Valley splitting in the heterostructure of MoTe$_2$/MnS$_2$ can reach up to 55.3 meV \cite{liang2024mechanisms}. At compressive strain, good valley splitting was reported in WSe$_2$/CrSnSe$_3$ \cite{khan2021strain}. For heterostructure MnPSe$_3$/WSe$_2$, more than 30 meV valley splitting was found \cite{D0TC03065A}. MoTe$_2$/WTe$_2$ showed giant valley splitting in a recent report \cite{zheng2016strain}. VX$_2$ (X = S, Se, Te) emerged as a member of ferrovalley materials \cite{fuh2016newtype}, where interaction of d electrons induced spontaneous valley splitting due to broken TRS and SIS.

Monolayer hexagonal VTe$_2$, VS$_2$, and VSe$_2$ are basic examples of ferrovalley materials. Room-temperature ferromagnetism was observed in ultrathin 3 nm VTe$_2$ grown by chemical vapor deposition (CVD), which provides its usefulness in valley polarization \cite{li2018synthesis}. Due to strong SOC effect and magnetism, giant valley splitting is observed in VTe$_2$ \cite{wang2021effects,musle2019temperature}. Additionally, it can be used in spin-calorimetry applications due to its higher spin polarizability \cite{musle2019temperature}. VSe$_2$ is useful for room temperature because of its high Curie temperature (up to 590 K) \cite{feng2018strain}. Recently, the synthesis of VSe$_2$ has encouraged researchers to use its potential as a magnetic and valleytronic material \cite{bonilla2018strong,fuh2016newtype}. It has both intrinsic valley splitting, ferromagnetism \cite{bonilla2018strong}, and possesses T and H phases \cite{pushkarev2019structural}. 
The h-VN, which also has a large Curie temperature of 768 K, can be used as a ferromagnetic half-metallic substrate for forming heterostructures \cite{kuklin2018two}. 

Spin-filtering and valleytronic properties in related systems have been explored in previous reports. For instance, 25\% W-doped monolayer VSe$_2$ (with SOC) was reported to filter spin-down electrons with an energy gap of 0.8~eV~\cite{Shihab2022ACSOmega}. Similarly, the 2H-VS$_2$/h-VN heterostructure exhibited strain-induced valley splitting, although with a relatively limited tunability~\cite{bian2022tunable}. Furthermore, the anomalous Hall conductivity (AHC) values obtained in several previous works~\cite{Ke} remain comparatively modest, leaving room for improvement in high-performance valleytronic and spintronic device applications.


Previous studies have examined spin-filtering and valleytronic behavior separately, but a combined realization in a single material has not yet been reported. Motivated by this gap, we systematically explored multiple stacking configurations of the previously unexamined VSe$_2$/VN heterostructure by varying the interlayer distance. We investigated the system’s stability using formation energy, phonon dispersion, and molecular dynamics simulations. After identifying the optimized heterostructure, we analyzed its electronic, magnetic, and structural properties, including band structure, density of states, magnetization easy axis, and Bader charge analysis using DFT calculations. Finally, by applying both compressive and tensile strain, we demonstrated a strain-tunable spin-filtering response together with giant valley splitting and AHC, unlocking the ability of this system for future spintronic and valleytronic device applications.

\section{Computational Details}

First-principles calculations of structural and electronic properties were performed using DFT-based tool Quantum ESPRESSO~\cite{giannozzi2009quantum}. The Perdew–Burke–Ernzerhof (PBE) functional was used for generalized gradient approximation (GGA) of exchange-correlation energy. Fully relativistic projector augmented wave (PAW) pseudopotentials of V, Se, and N were used to account for the effect of SOC. The DFT-D3 method was used for van der Waals correction. The plane-wave kinetic energy cutoff was selected as 680 eV for expanding the wave functions, and the cutoff was selected as 6800 eV for charge density and potentials. Structural relaxation was performed until forces on atoms had converged down to $10^{-2}$ eV/\AA \, and total energy converged below the threshold of $10^{-6}$ eV/\AA. Due to the strong repulsion of localized d electrons, the GGA+U method was used with U = 2.00 eV and J = 0.87 eV for the 3d orbitals of V~\cite{fuh2016newtype}. A Gamma-centered Monkhorst-Pack scheme with a $k$-point mesh of ($8 \times 8 \times 1$) and ($12 \times 12 \times 1$) was used for the relaxation and self-consistent field (SCF) calculations, respectively. To obtain the density of states (DOS) and the projected density of states (PDOS), a $k$-grid of ($32 \times 32 \times 1$) was used in non-self-consistent field (NSCF) calculations. To estimate the accurate band structure the Heyd-Scuseria–Ernzerhof (HSE06) was used. 

E-$k$ diagrams were constructed along the path $\mathrm{M} \to \mathrm{K} \to \mathrm{\Gamma} \to \mathrm{K}' \to \mathrm{M}$ of the first Brillouin zone, with 35 $k$-points between each high-symmetry point. To avoid interactions between adjacent layers, a vacuum layer with a thickness of 20 \AA\ was added. The valley splitting was defined as $\mathrm{\Delta C_{KK'}}$ for the conduction band (CB) and 
$\mathrm{\Delta V_{KK'}}$ for the valence band (VB), where both 
$\mathrm{\Delta C_{KK'}}$ and $\mathrm{\Delta V_{KK'}}$ are given by 
$\mathrm{E_{K} - E_{K'}}$. Here, $\mathrm{E_{K}}$ and $\mathrm{E_{K'}}$ denote 
the energy (top of the VB and bottom of the CB) at the $\mathrm{K}$ and $\mathrm{K'}$ points of the Brillouin zone, respectively.

Moreover, projected band structures were calculated to show the contribution of orbitals of a particular atom in the band structure. Biaxial strain ($\varepsilon$) was calculated using $\varepsilon = \frac{a - a_0}{a_0}$, where $a$ is the strained lattice constant and $a_0$ is the equilibrium lattice constant. Spin projection on band structure was shown using a color map by varying the spin operator from -0.50 to 0.50.

To observe the dynamic stability of the heterostructure, the Phonopy package ~\cite{phonopy-phono3py-JPCM} was used. Here, the dimension of the supercell was taken as ($4 \times 4 \times 1$) and a large value of kinetic energy cutoff (950 eV) was used. To avoid flexural phonons near $\Gamma$ point, the structure was assumed to be isolated as 2D. To plot the phonon bands, $\mathrm{\Gamma} \to \mathrm{M} \to \mathrm{K} \to \mathrm{\Gamma}$ path was selected with 100 band points. The \textit{ab initio} molecular dynamics (AIMD) simulations were performed in the canonical (NVT) ensemble using the Nos\'e--Hoover thermostat for 5~ps to assess the thermal stability of the system. For energetic stability, the formation energy was calculated for different stackings of the heterostructure by,
\vspace{-0.3cm}
\begin{equation}
\mathrm{E}_{\mathrm{formation}} = \mathrm{E}_{\mathrm{VSe_2/VN}} - \mathrm{E}_{\mathrm{VSe_2}} - \mathrm{E}_{\mathrm{VN}}
\end{equation}

Charge density difference (CDD) was calculated using the formula, 
\begin{equation}
\Delta \rho(\mathbf{r}) = \rho_{\text{VSe}_2/\text{VN}}(\mathbf{r}) - \rho_{\text{VSe}_2}(\mathbf{r}) - \rho_{\text{VN}}(\mathbf{r})
\end{equation}
where $\rho_{\text{VSe}_2/\text{VN}}(\mathbf{r})$ is the charge density of the full heterostructure, $\rho_{\text{VSe}_2}(\mathbf{r})$ is the charge density of the isolated VSe$_2$ monolayer (in the same atomic positions as in the heterostructure), and $\rho_{\text{VN}}(\mathbf{r})$ is the charge density of the isolated VN monolayer (in the same atomic positions as in the heterostructure). Here, $\mathbf{r}$ denotes the space vector. Bader charge analysis was conducted using the code by Henkelman group \cite{HENKELMAN2006354}. Work function was calculated using the formula,
\vspace{-0.3cm}
\begin{equation}
\hspace{2cm}\phi = \mathrm{E}_{\mathrm{vacuum}} - \mathrm{E_F}
\end{equation}
where $\mathrm{E}_{\mathrm{vacuum}}$ is the vacuum energy level and $\mathrm{E_F}$ is the Fermi level. Spin density was calculated using:
\begin{equation}
\hspace{1cm} \Delta \rho_{\mathrm{s}}(\mathbf{r}) \;=\; \rho_{\uparrow}(\mathbf{r}) \;-\; \rho_{\downarrow}(\mathbf{r}) 
\end{equation}

Here, $\rho_{\uparrow}$($\mathbf{r}$) denotes the positive spin density and $\rho_{\downarrow}$($\mathbf{r}$) denotes the negative spin density.

To calculate Berry curvature and anomalous Hall conductivity, the maximally localized Wannier functions (MLWF) were constructed using Wannier90 \cite{Pizzi2020} with 60 Wannier functions generated from 60 Bloch bands spanning the relevant energy window. Projections included the s, p and d orbitals of V atoms and s and p orbitals of Se and N atoms. The initial Wannierization was performed on a Monkhorst--Pack (12 $\times$ 12 $\times$ 1) $k$-point grid, from which the real-space Hamiltonian and tight-binding model were obtained. Berry curvature calculations were carried out using the Wannier90 post-processing tool with respect to the path $\mathrm{M} \to \mathrm{K} \to \mathrm{\Gamma} \to \mathrm{K}' \to \mathrm{M}$. The AHC was computed using the Berry curvature on an interpolated $k$-point mesh of (100 $\times$ 100 $\times$ 1).

\section{Results and Discussion}
\subsection{Pristine Hexagonal VSe$_2$ and VN Monolayer}
\begin{figure*}[h]
\centering
\includegraphics[scale=0.29]{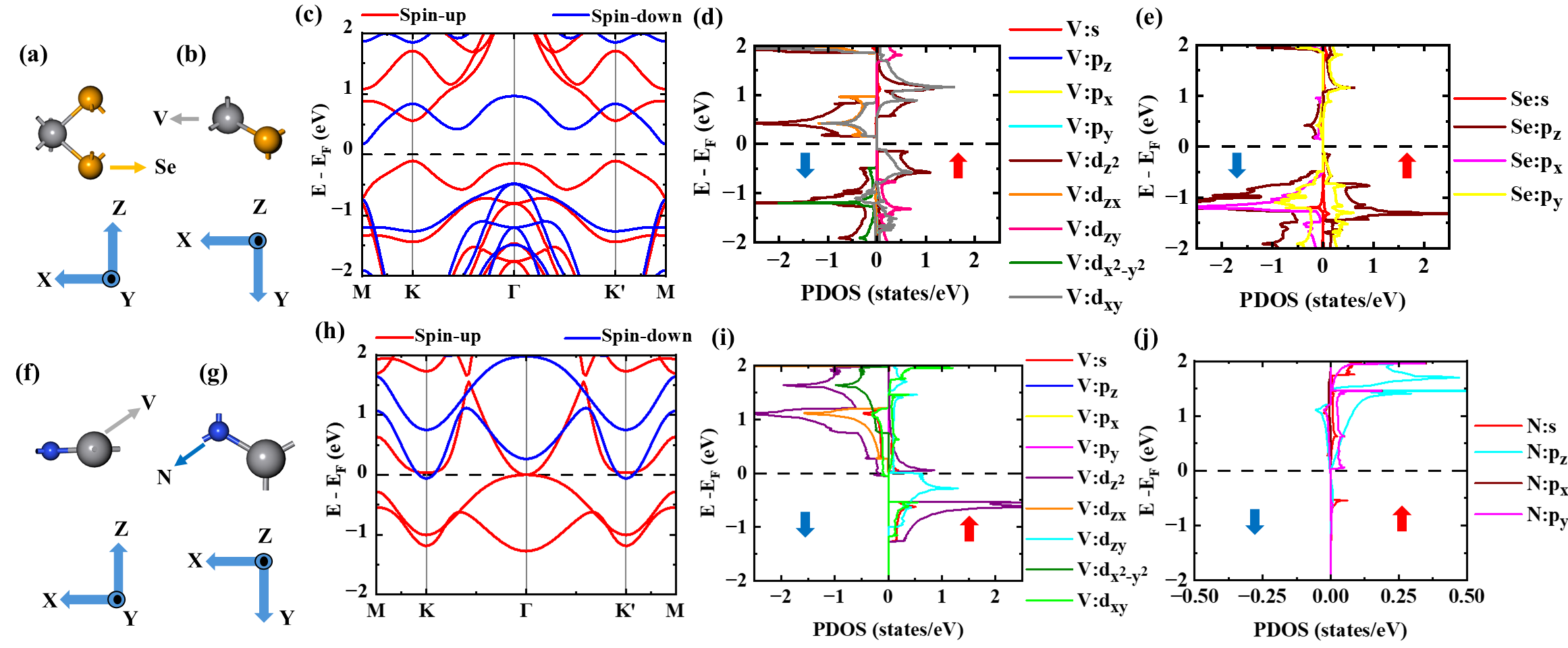}
  \caption{(a) Side view of pristine hexagonal VSe$_2$ monolayer. (b) Top view of pristine hexagonal VSe$_2$ monolayer. (c) Spin-polarized band structure (without SOC) of pristine hexagonal VSe$_2$ monolayer. (d) PDOS of $\textnormal{s}$, $\textnormal{p}$ and $\textnormal{d}$ orbitals of the V atom. (e) PDOS of $\textnormal{s}$, $\textnormal{p}$ orbitals of the Se atom. (f) Side view of pristine hexagonal VN monolayer. (g) Top view of pristine hexagonal VN monolayer. (h) Spin-polarized band structure (without SOC) of pristine hexagonal VN monolayer. (i) PDOS of $\textnormal{s}$, $\textnormal{p}$ and $\textnormal{d}$ orbitals of the V atom. (j) PDOS of $\textnormal{s}$, $\textnormal{p}$ orbitals of the N atom. The red line and red arrow represent the spin-up channel, while the blue line and blue arrow represent the spin-down channel.}
  \label{pristinevse2andvn}
\end{figure*}

The structural model, electronic band structure, and PDOS of the pristine hexagonal VSe$_2$ monolayer are shown in Figs.~\ref{pristinevse2andvn}(a–e). The side and top views appear in Figs.~\ref{pristinevse2andvn}(a) and (b), respectively. The unit cell of this 2D magnetic semiconductor comprised three atoms: one V and two Se; and as summarized in Table~\ref{tab:vse2-vn-params}, was characterized by a lattice parameter of 3.347~\AA, a V–Se bond length of 2.50~\AA, an induced magnetization of \(1.00~\mu_{\mathrm{B}}\) per unit cell and a Se–V–Se bond angle of \(79.3^\circ\). These structural parameters were consistent with values reported in the literature~\cite{Shihab2022ACSOmega}. Fig.~\ref{pristinevse2andvn}(c) displays the spin-polarized band structure (without SOC) of monolayer VSe$_2$, where the valence band maximum (VBM) occurred at $\mathrm{K}'$ and the conduction band minimum (CBM) occurred at M, revealing an indirect band gap. Figs. S1(a) and (b) show the spin-polarized band structure (with SOC) together with the work function. We found that the valley splitting \(\Delta \mathrm{V}_{\mathrm{KK'}}\), bandgap, and work function were 79.2~meV, 0.29~eV and 5.47~eV respectively, in agreement with previous studies~\cite{feng2018strain,wasey2015quantum}.

The PDOS analysis, shown in Figs.~\ref{pristinevse2andvn}(d) and (e), highlights the contributions of different atomic orbitals to the electronic structure of VSe$_2$. Fig.~\ref{pristinevse2andvn}(d) presented the orbital PDOS for V:\,$\textnormal{s}$, V:\,$\textnormal{p}$, V:\,$\textnormal{d}$ states, and the V:$\textnormal{d}$ orbitals dominated near the Fermi level ($\mathrm{E_F}$). At the CBM of VSe$_2$, the contributions primarily arose from V:\,$\textnormal{d}$ orbitals; especially $\textnormal{d}_{z^2}$, $\textnormal{d}_{zx}$, and $\textnormal{d}_{xy}$ in the spin-down channel. In contrast, the VBM of VSe$_2$ was contributed by V:\,$\textnormal{d}$ orbitals, mainly $\textnormal{d}_{z^2}$ and $\textnormal{d}_{zx}$, in the spin-up channel. Fig.~\ref{pristinevse2andvn}(e) shows the PDOS from Se:\,$\textnormal{s}$ and Se:\,$\textnormal{p}$ orbitals. At both VBM and CBM, Se contributions were very minor. By contrast, Se exhibited larger contributions than V in the lower valence bands for both spin channels; most notably Se:\,$\textnormal{p}_{z}$, Se:\,$\textnormal{p}_{x}$, and Se:\,$\textnormal{p}_{y}$ in the spin-down channel and Se:\,$\textnormal{p}_{z}$, Se:\,$\textnormal{p}_{y}$ in the spin-up channel.

\begin{table}[t]
  \centering
  \caption{Structural, electronic, and magnetic properties of pristine VSe$_2$ and VN monolayers}
  \label{tab:vse2-vn-params}
  \begin{tabular}{lcc}
    \toprule
    \textbf{Property} & \textbf{VSe$_2$} & \textbf{VN} \\
    \midrule
    Lattice parameter (\AA)          & 3.347 & 3.224 \\
    Bond length (\AA)                & 2.50  & 1.86  \\
    Bond angle             & ~~79.3 $^{\circ}$  & ~~120 $^{\circ}$   \\
    Magnetization ($\mu_\mathrm{B}$ per unit cell) & 1.00  & 1.98  \\
    Work function (eV)               & 5.47  & 3.81  \\
    \bottomrule
  \end{tabular}
\end{table}

Figs.~\ref{pristinevse2andvn}(f--j) present the structural models, electronic band structures, and projected density of states (PDOS) for a pristine hexagonal VN monolayer. The side and top views are shown in Figs.~\ref{pristinevse2andvn}(f) and (g). As referred (Table~\ref{tab:vse2-vn-params}), we calculated the values of lattice parameter, bond length, bond angle, and induced magnetization after optimization, which were 3.224~\AA, 1.86~\AA, \(120^\circ\), and \(1.98~\mu_{\mathrm{B}}\) per unit cell, respectively. The spin-polarized band structure (without SOC) in Fig.~\ref{pristinevse2andvn}(h) revealed a half-metallic ground state, while the corresponding calculation (with SOC) was provided in Fig.~S2(a). These results are consistent with previous reports~\cite{hua2025first}. We determined the work function of the hexagonal VN monolayer to be 3.81~eV (Fig.~S2(b)), which is also in close agreement with earlier findings~\cite{bian2022tunable}.

The orbital-resolved PDOS of V and N atoms are displayed in Figs.~\ref{pristinevse2andvn}(i) and (j). The PDOS was dominated by the V atom in the pristine hexagonal VN monolayer. The N atom contributed less to the PDOS, consistent with reduced orbital overlap associated with its smaller size. Among the N orbitals, the \(\textnormal{p}_{z}\) orbital in the spin-up channel contributed the most, although it remained much smaller than the V contribution; in the spin-down channel, the N contribution was negligible. By contrast, the V atom exhibited prominent weight in both spin channels. A notable feature is that the spin-down channel showed no states below \(\mathrm{E_F}\); near and above \(\mathrm{E_F}\), V:\,\(\textnormal{d}_{z^2}\) and V:\,\(\textnormal{d}_{zx}\) contributed strongly, with a smaller contribution from V:\,\(\textnormal{d}_{xy}\). In the spin-up channel, finite weight appeared both below and above \(\mathrm{E_F}\), with V:\,\(\textnormal{d}_{z^2}\) and V:\,\(\textnormal{d}_{zy}\) providing the largest contributions.


\subsection{Structural and Stability Analysis of VSe$_2$/VN Heterostructure}

Both VSe$_2$ and VN monolayers have been synthesized experimentally. Prior studies~\cite{liu2018epitaxially,you2022salt} demonstrate that VSe$_2$ can be grown epitaxially by molecular beam epitaxy (MBE) and by chemical vapor deposition (CVD). Thin VN films have also been realized using unfiltered cathodic arc evaporation (CAE)~\cite{warcholinski2024mechanical}. In addition, \textit{ab initio} calculations support the dynamical stability of these monolayers~\cite{Shihab2022ACSOmega,hua2025first}. As summarized in Table~\ref{tab:vse2-vn-params}, we evaluated the lattice mismatch between VSe$_2$ and VN monolayers was \(3.82\%\), which is below the commonly used \(5\%\) compatibility threshold~\cite{chen2019tunable}, making this pair a promising candidate for heterostructure fabrication.

\begin{figure*}[h]
\centering
\includegraphics[width=\textwidth]{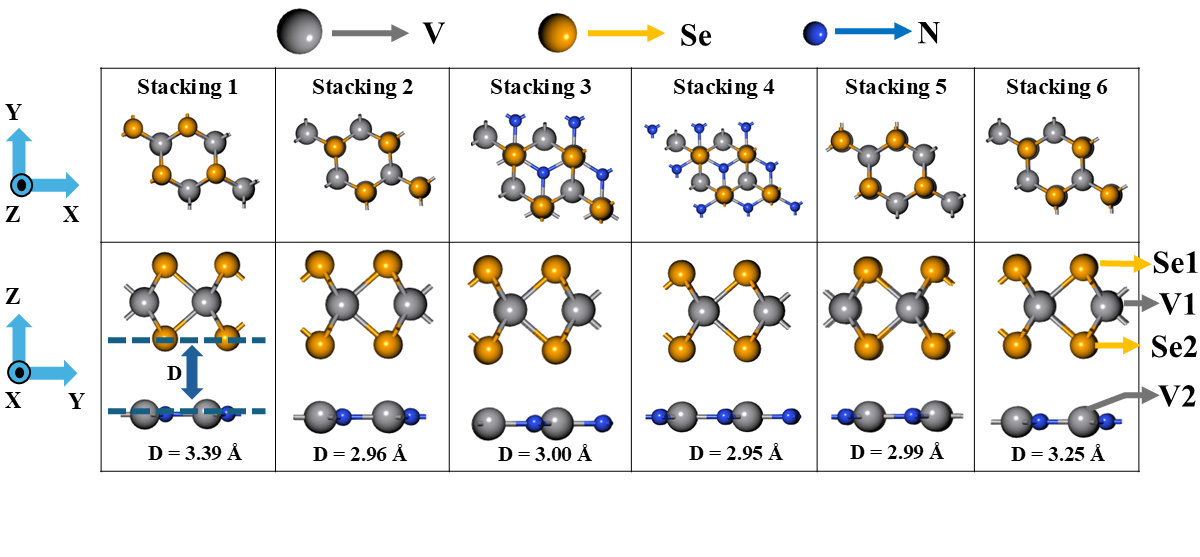}
  \caption{Different stackings of VSe$_2$/VN ($2 \times 2 \times 1$ supercell) heterostructure. First row indicates the top views and second row indicates the side views of the heterostructure with varying interlayer distance (D). }
  \label{models}
\end{figure*}

Fig.~\ref{models} illustrates the various stacking patterns in the VSe$_2$/VN heterostructure. The top row displays top views of six configurations and the bottom row shows side views with the corresponding interlayer distance (D). Configurations 1, 2, 5, and 6 followed an AA stacking, whereas configurations 3 and 4 adopted AB stacking. From Table~\ref{tab:stacking-summary}, we determined that only stackings 1 and 2 had negative formation energies; the others were energetically unfavorable. Among these, stacking~2 emerged as the most stable, with a formation energy of \(-0.95\)~eV per unit cell and an in-plane lattice parameter of \(3.328\)~\AA\ per unit cell. This level of stabilization is comparable to that reported for other 2D van der Waals heterostructures, such as MoS$_2$/ZnO~\cite{liu2024studies}.

\begin{figure*}[h]
\centering
\includegraphics[width=0.8\textwidth]{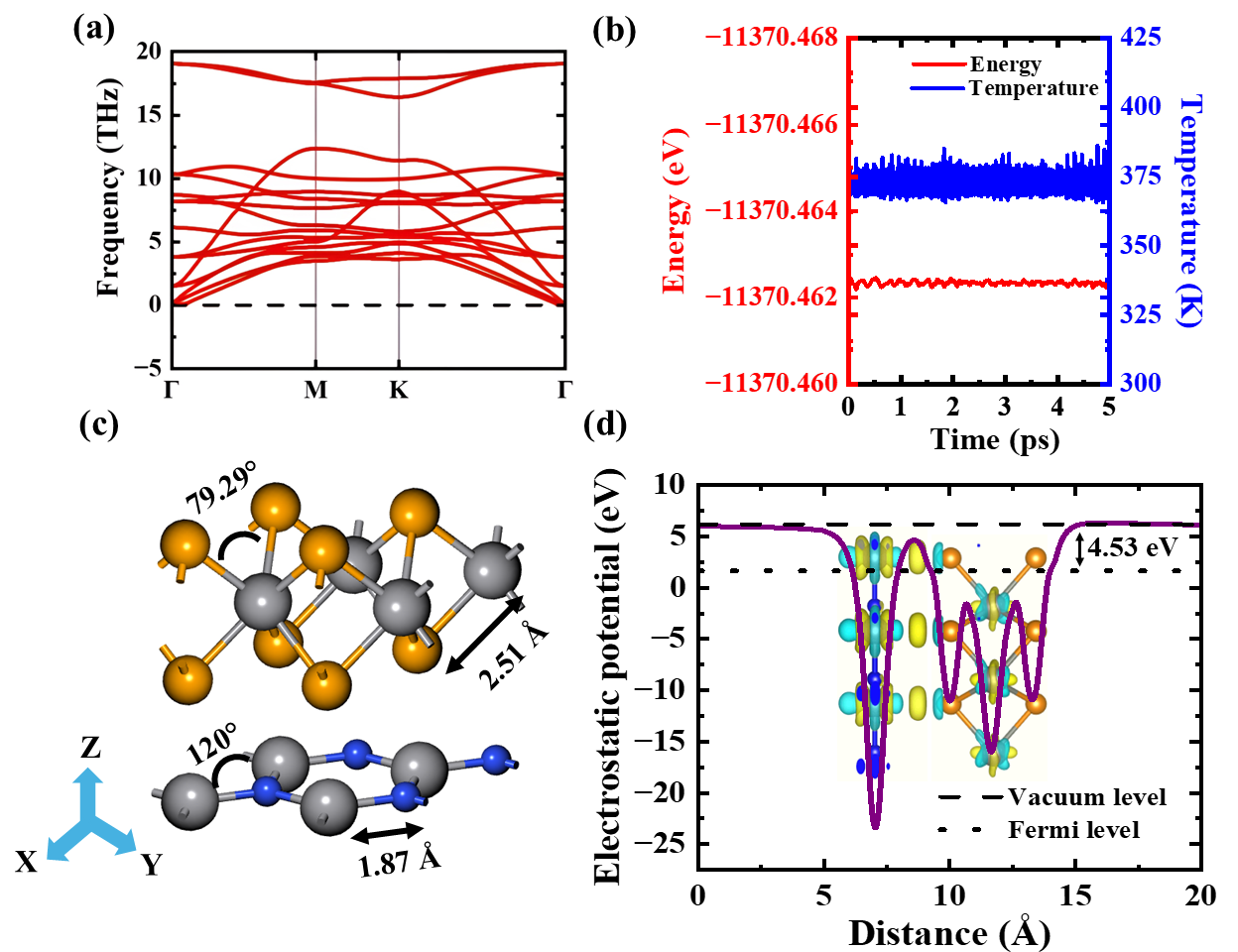}
  \caption{(a) Phonon dispersion diagram of the VSe$_2$/VN heterostructure showing the absence of any imaginary frequencies, confirming its dynamic stability. (b) Variation of total energy and temperature of the VSe$_2$/VN heterostructure under the NVT ensemble at 373 K, recorded from 0 to 5 ps with a timestep of 1 fs. (c) Optimized atomic structure of the VSe$_2$/VN heterostructure. (d) Work function along with CDD diagram (with isosurface value of $0.0033~\mathrm{e}\,\AA^{-3}$)
 of the VSe$_2$/VN heterostructure with respect to z-axis, where the dashed line represents the vacuum level and the dotted line denotes the Fermi level. Cyan and yellow represent electron depletion and accumulation, respectively.
 }
  \label{structstability}
\end{figure*}

To assess the dynamical and thermal stability of VSe$_2$/VN, we computed the phonon dispersion and performed AIMD, as shown in Figs.~\ref{structstability}(a) and (b). In Fig.~\ref{structstability}(a), the phonon spectrum contains 15 branches; three times the number of atoms in the unit cell (five) as expected. The absence of imaginary frequencies ensured the system’s dynamical stability. Figure~\ref{structstability}(b) presents the total energy evolution at 373~K over 5~ps using the NVT ensemble. The only minor fluctuations observed in the total energy indicate that the VSe$_2$/VN heterostructure is thermally stable under these conditions.

\begin{table}[t]
\centering
\caption{Structural properties and formation energies for the six stacking configurations}
\label{tab:stacking-summary}
\resizebox{\columnwidth}{!}{%
\begin{tabular}{lccc}
\toprule
\textbf{Stacking} &
\makecell{\textbf{Interlayer distance}\\\textbf{(Å)}} &
\makecell{\textbf{Lattice parameter}\\\textbf{(Å / unit cell)}} &
\makecell{\textbf{Formation energy}\\\textbf{(eV / unit cell)}} \\
\midrule
Stacking 1 & 3.39 & 3.230 & $-0.53$ \\
Stacking 2 & 2.96 & 3.328 & $-0.95$ \\
Stacking 3 & 3.00 & 3.256 & $~~0.26$ \\
Stacking 4 & 2.95 & 3.257 & $~~0.21$ \\
Stacking 5 & 2.99 & 3.251 & $~~0.22$ \\
Stacking 6 & 3.25 & 3.250 & $~~0.41$ \\
\bottomrule
\end{tabular}}
\end{table}

\begin{table}[htbp]
\centering
\caption{Optimized structural, electronic, and magnetic properties of the VSe$_2$/VN heterostructure}
\label{tab:vse2vn_params}
\begin{tabular}{lc}
\hline
\textbf{Property} & \textbf{Value} \\
\hline
Lattice parameter (\AA) & 3.328 \\ 

Bond angle  &
\begin{tabular}[t]{@{}c@{}}
Se--V--Se: 79.29 $^{\circ}$\\
V--N--V / N--V--N: 120 $^{\circ}$
\end{tabular} \\

Bond length (\AA) &
\begin{tabular}[t]{@{}c@{}}
V--Se: 2.51 \\
V--N: 1.87
\end{tabular} \\

Magnetization ($\mu_{\mathrm B}$ per unit cell) & 3.00 \\
Work function (eV) & 4.53 \\
Interlayer distance (\AA) & 2.96 \\
\hline
\end{tabular}
\end{table}

Fig.~\ref{structstability}(c) shows the optimized structural parameters of the VSe$_2$/VN heterostructure. As depicted in Table~\ref{tab:vse2vn_params}, We found the unit cell lattice parameter, Se--V--Se and V--N--V/N--V--N bond angles to be 3.328~\AA, 79.29$^{\circ}$, and 120$^{\circ}$, respectively. We determined the V--Se bond length to be 2.51~\AA~and the V--N bond length to be 1.87~\AA. The total magnetization was found to be 3.00~$\mu_{\mathrm{B}}$ per unit cell. We evaluated the work function of the VSe$_2$/VN heterostructure to be 4.53~eV (Fig.~\ref{structstability}(d)), obtained from the difference between the vacuum level and the Fermi level. The interlayer distance between the monolayers was 2.96~\AA, as mentioned for stacking~2.


\subsection{Electronic Property of VSe$_2$/VN Heterostructure}

We employed Bader charge analysis to quantify the net charge transfer for each atom in the VSe$_2$/VN heterostructure. In this approach, the atomic charge $Q_i$ was evaluated from the Bader electron count and the valence reference using the formula below:

\begin{equation}
 \hspace{2cm}   Q_i \equiv Z_i^{\mathrm{val}} - q_i^{\mathrm{B}}
    \label{eq:atomic_charge}
\end{equation}
where $Q_i$ is the net atomic charge on atom $i$, $Z_i^{\mathrm{val}}$ is its valence electron count and $q_i^{\mathrm{B}}$ is the Bader charge obtained from the charge density analysis.

\begin{table}[t]
  \centering
  \caption{Bader charge analysis (in e) for isolated monolayers and the heterostructure and the resulting net charge transfer, $\Delta N$ (heterostructure $-$ isolated). Negative $\Delta N$ indicates electron gain and positive $\Delta N$ indicates electron loss }
  \label{tab:bader-charges}
  \begin{tabular}{lccc}
    \toprule
    \textbf{Element} & \textbf{Isolated layer} & \textbf{Heterostructure} & \textbf{$\Delta N$} \\
    \midrule
    V1  & +1.18 & +1.15 & \phantom{+}--0.03 \\
    Se1 & --0.59 & --0.59 & \phantom{+}~0.00 \\
    Se2 & --0.59 & --0.67 & \phantom{+}--0.08 \\
    V2  & +1.49 & +1.52 & ~+0.03 \\
    N   & --1.49 & --1.41 & ~+0.08 \\
    \bottomrule
  \end{tabular}
\end{table}

In Table~\ref{tab:bader-charges}, we computed \(Q_i\) for V1 (V atom in VSe\(_2\)), V2 (V atom in VN), Se1 (Se atom far from VN), Se2 (Se atom close to VN), and the N atom for both the isolated monolayers and the heterostructure, and then evaluated the net charge transfer, \(\Delta N\) (heterostructure \(-\) isolated). For isolated VSe\(_2\), the V atom showed electron loss \((+1.18~\mathrm{e})\) and each Se atom showed electron accumulation \((-0.59~\mathrm{e})\). In isolated VN, the V atom likewise exhibited electron loss \((+1.49~\mathrm{e})\), larger than for V1, while the N atom exhibited electron accumulation \((-1.49~\mathrm{e})\), maintaining charge neutrality.

After forming the heterostructure, charge redistributed and the individual monolayers were no longer neutral. The calculated \(\Delta N\) values indicated electron gain for V1 \((-0.03~\mathrm{e})\) and Se2 \((-0.08~\mathrm{e})\), electron depletion for V2 \((+0.03~\mathrm{e})\) and N \((+0.08~\mathrm{e})\), and essentially no net transfer for Se1 \((0.00~\mathrm{e})\). We therefore concluded that electrons depleted from the VN layer and accumulated on the VSe\(_2\) layer upon heterostructure formation.

This conclusion was supported by the CDD diagram and the work function analysis in Fig.~\ref{structstability}(d). The CDD showed electron accumulation near the VSe\(_2\) layer and depletion near the VN layer, with no change at Se1. Consistently, the work functions of VSe\(_2\), VN, and the VSe\(_2\)/VN heterostructure were \(5.47~\mathrm{eV}\), \(3.81~\mathrm{eV}\), and \(4.53~\mathrm{eV}\), respectively (Tables~\ref{tab:vse2-vn-params} and~\ref{tab:vse2vn_params}). Because electrons flow from the lower to the higher work function until the Fermi levels are aligned, charge was transferred from VN to VSe\(_2\), stabilizing the heterostructure at a work function of \(4.53~\mathrm{eV}\).

\begin{figure*}[h]
\centering
\includegraphics[width=\textwidth]{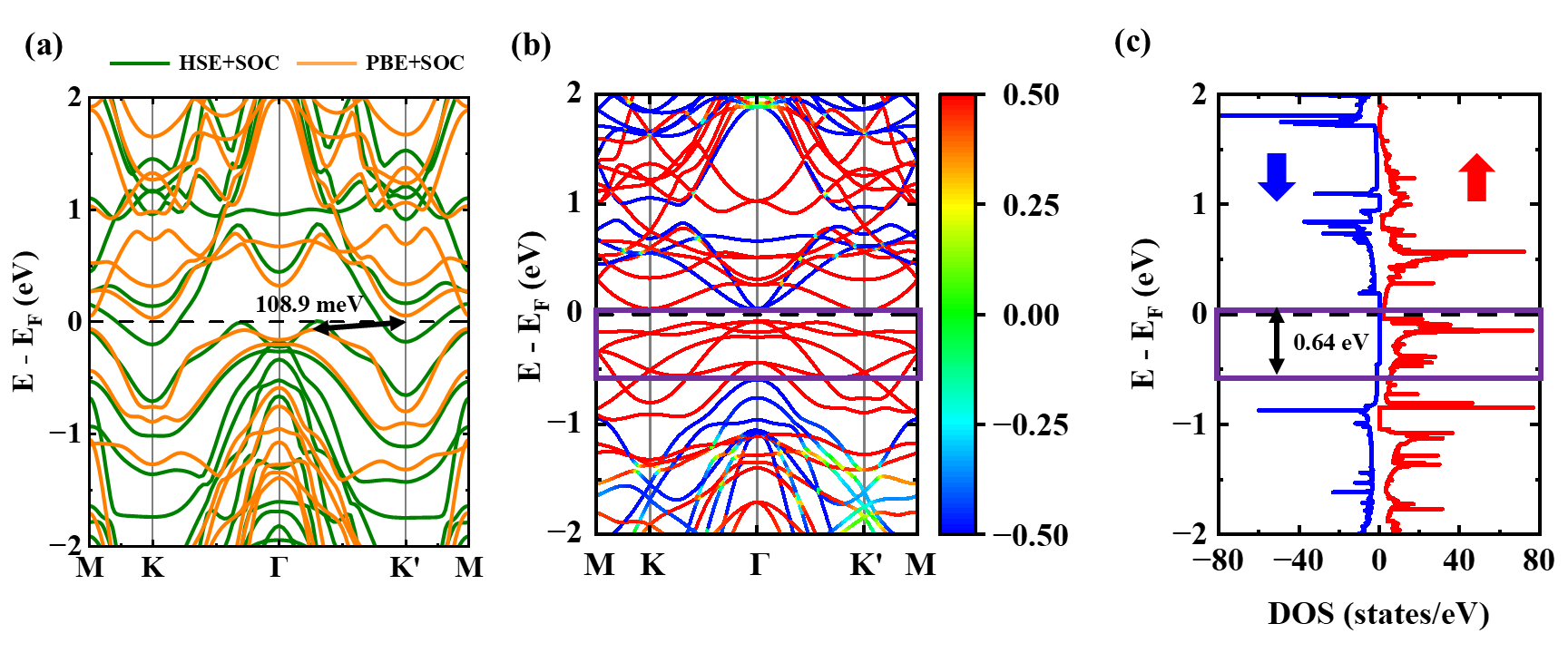}
  \caption{(a) Band structure (with SOC) of unit cell VSe$_2$/VN heterostructure. Green line indicates band structure with HSE06 functional and orange line indicates band structure with PBE functional. (b) Band structure (with SOC) ($2 \times 2 \times 1$ supercell) of VSe$_2$/VN heterostructure. The color map indicates expectation values of the spin operator on the spinor wave-functions ranging from -0.50 (blue) to +0.50 (red). (c) Spin-polarized DOS ($2 \times 2 \times 1$ supercell) of VSe$_2$/VN heterostructure. Blue arrow, red arrow indicate spin-down channel and spin-up channel, respectively. The purple box shows the forbidden energy window (0.64 eV) of spin-down channel in (b) and (c). }
  \label{bsdos}
\end{figure*}


\begin{figure*}[h]
\centering
\includegraphics[width=\textwidth]{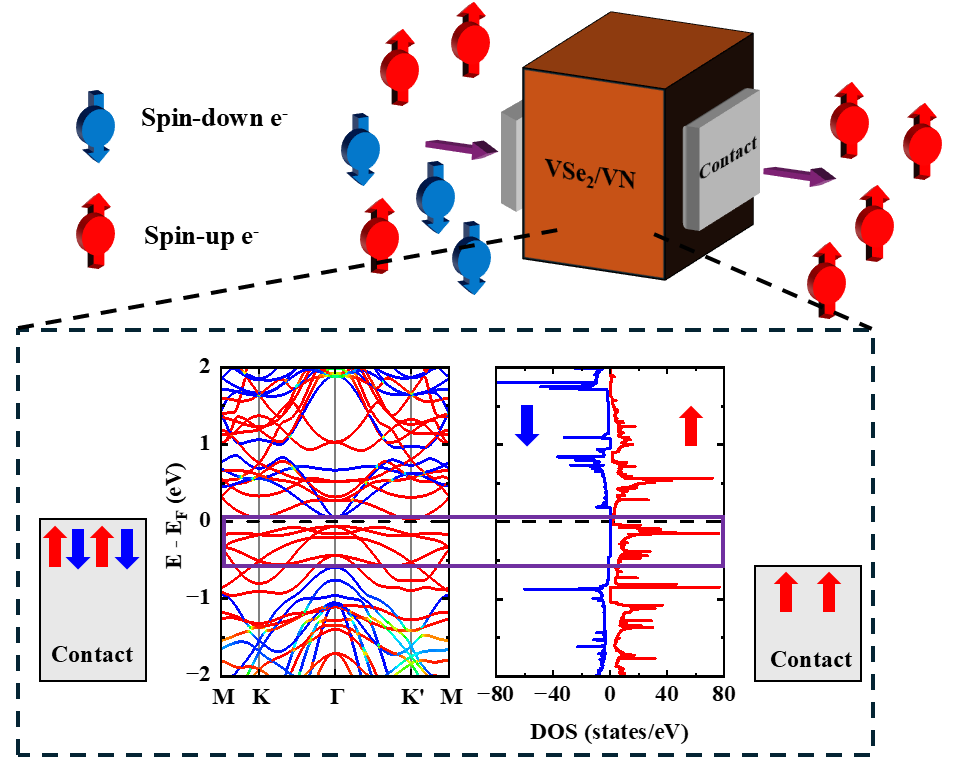}
  \caption{Schematic of a spin-filter using the VSe$_2$/VN ($2 \times 2 \times 1$ supercell) heterostructure. By applying external perturbation, spin-up electrons will be available for a certain energy window at the right contact.}
  \label{spinfilter}
\end{figure*}


We considered SOC in all band structure calculations to expose valley splitting. For a 2D single layer with two inequivalent valleys labeled by $\tau=\pm1$ (for $\mathrm{K}$ and $\mathrm{K}'$), a two-band $k\cdot p$ basis $\{|c\rangle,|v\rangle\}$ near each valley can be defined where Pauli matrices $\sigma_{x,y,z}$ act in the $(c,v)$ subspace and a spin index $s=\pm1$ denotes $\uparrow,\downarrow$. The spinless Dirac-type block can be defined~\cite{xiao2012coupled}:
\begin{equation}
H_0(\mathbf{k},\tau)=\hbar v\big(\tau k_x\sigma_x+k_y\sigma_y\big)+\frac{\Delta}{2}\sigma_z 
\end{equation}
Here, $v$ is the effective Dirac velocity, $\Delta$ is the spinless band gap. SOC at the band edges is captured by valley and spin-dependent shifts that projects onto the conduction and valence sectors:
\begin{equation}
H_{\mathrm{SOC}}(\tau,s)=\tau s\big(\lambda_c\Pi_c+\lambda_v\Pi_v\big), \qquad \Pi_{c/v}=\frac{\sigma_0\pm\sigma_z}{2}
\end{equation}
where $\sigma_0$ is the identity in band space,  $\Pi_{c/v}$ projects onto the conduction/valence edges, $\tau=\pm1$ labels the valleys, $s=\pm1$ labels the out-of-plane spin, and $\lambda_{c/v}$ are the SOC coupling strengths at the conduction/valence edges. At the band extrema ($\mathbf{k}=0$), $H_0+H_{\mathrm{SOC}}$ is diagonal, yielding
\begin{equation}
\mathrm{E_c}(\tau,s)=+\frac{\Delta}{2}+\tau s\,\lambda_c, \qquad \mathrm{E_v}(\tau,s)=-\frac{\Delta}{2}-\tau s\,\lambda_v 
\end{equation}
Fixing the spin $s$, the valley splittings (energy differences between $\mathrm{K}$ and $\mathrm{K}'$) are
\begin{equation}
\mathrm{\Delta C_{KK'}}=\mathrm{E_c}(+1,s)-\mathrm{E_c}(-1,s)=2\lambda_c s
\end{equation}
\begin{equation}
\mathrm{\Delta V_{KK'}}=\mathrm{E_v}(+1,s)-\mathrm{E_v}(-1,s)=2\lambda_v s
\end{equation}
 Thus, SOC introduces valley splitting, which is also applicable for 2D heterostructures.

Figs.~\ref{bsdos}(a-c) represent the band structure and DOS of the VSe$_2$/VN heterostructure. In Fig.~\ref{bsdos}(a), the band structures (including SOC) are obtained using the PBE and HSE06 functionals. A small indirect band gap of $108.9$~meV was found, extending from $\mathrm{K}'$ to a $k$-point between $\Gamma$ and $\mathrm{K}'$, using the PBE functional, whereas the HSE06 functional resulted in a half-metallic nature of the heterostructure. The conduction band valley splitting was found to be more dominant. Using PBE and HSE06, $\mathrm{\Delta C_{KK'}}$ was found to be $22.9$~meV and $23.7$~meV in the unit cell, respectively. As HSE06 is computationally expensive and the valley splittings were close for PBE and HSE06, further calculations were performed using the PBE functional. We calculated the PBE spin-polarized band structure ($2\times2\times1$ supercell) of the heterostructure to account for the band-folding effect, as shown in Fig.~\ref{bsdos}(b). It was found that the spin-down channel showed a semiconducting gap ($0.64$~eV), whereas the spin-up channel was nearly gapless (approximately $108.9$~meV), consistent with the PBE result for the unit cell. The spin-polarized DOS of this supercell was also found to be consistent with the corresponding band structure; the spin-down channel exhibited a gap, whereas the spin-up channel was nearly gapless, giving the system a half-metallic (close to spin-gapless) character.

This idea can be used to form a spin-filtering device, as mentioned in the study~\cite{Shihab2022ACSOmega}. As illustrated in Fig.~\ref{spinfilter}, the $0.64$~eV energy window provides sufficient available states for the spin-up channel, whereas no states are present for the spin-down channel. Therefore, within this window, by applying external perturbation, spin-up electrons can be selectively transmitted and filtered out at the right contact.

\begin{figure*}[h]
\centering
\includegraphics[width=0.8\textwidth]{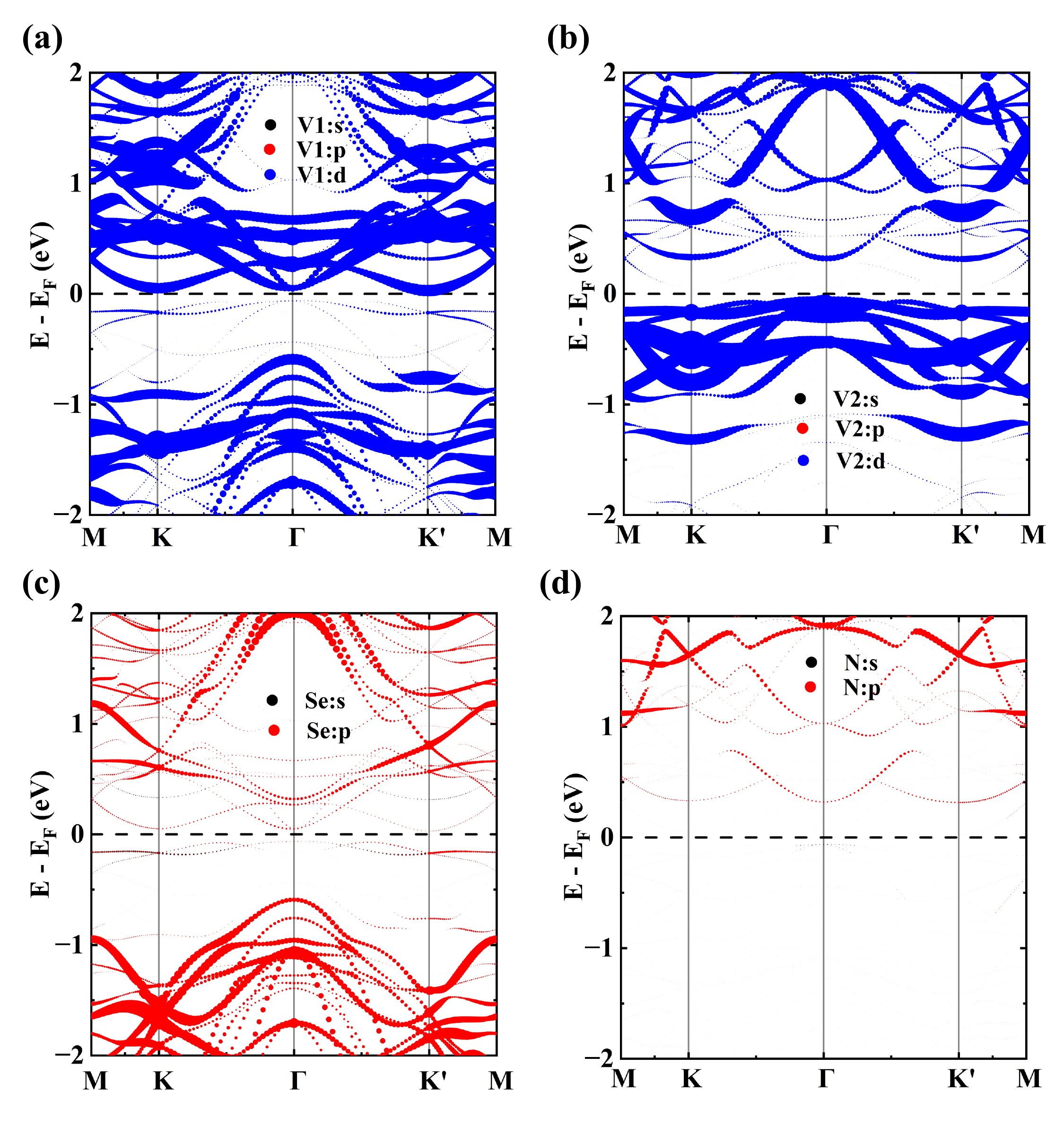}
  \caption{Orbital projected band structure of VSe$_2$/VN ($2 \times 2 \times 1$ supercell). (a) Projection of  s, p, d orbitals of V1 (V atom of VSe$_2$). (b) Projection of s, p, d orbitals of V2 (V atom of VN).  (c) Projection of s, p orbitals of Se. (d) Projection of s, p orbitals of N.}
  \label{fat}
\end{figure*}

In Figs.~\ref{fat}(a-d), the orbital projected band structure of VSe$_2$/VN are shown. In Fig.~\ref{fat}(a), V1:\,$\textnormal{d}$ contributed most strongly above $\mathrm{E_F}$, whereas in Fig.~\ref{fat}(b) V2:\,$\textnormal{d}$ exhibited the larger weight below $\mathrm{E_F}$. The V:\,$\textnormal{s}$ and V:\,$\textnormal{p}$ orbitals contributed negligibly. In contrast, as shown in Fig.~\ref{fat}(c), Se:\,$\textnormal{p}$ carried noticeable weight at lower-lying (deeper) energies but contributed negligibly near $\mathrm{E_F}$. The N:\,$\textnormal{p}$ contribution was comparatively weak, consistent with the smaller size of the N atom (Fig.~\ref{fat}(d)). Similar to V, the Se:\,$\textnormal{s}$ and N:\,$\textnormal{s}$ orbitals contributed negligibly across the energy range.

Figs.~S4(a-d) illustrate the PDOS (without SOC) to provide a clearer picture of the orbital contributions. Table~\ref{tab:orbital-contrib-transposed} summarizes the orbital-decomposed weights for VSe$_2$/VN. Near $\mathrm{E_F}$ in the spin-up channel, the spectrum was dominated by V2:\,$\textnormal{d}_{z^2}$ (42.37\%) and Se:\,$\textnormal{p}_{z}$ (25.44\%), with additional weight from V1:\,$\textnormal{d}_{z^2}$ (17.11\%), V2:\,$\textnormal{d}_{xy}$ (10.01\%), and a minor Se:\,$\textnormal{p}_{y}$ component (5.07\%); Se:\,$\textnormal{p}_{x}$ was negligible. For the spin-down channel at the CBM, the character was V:\,$\textnormal{d}$ dominated, led by V1:\,$\textnormal{d}_{z^2}$ (51.70\%) with sizable V1:\,$\textnormal{d}_{zx}$ (19.58\%), and V1:\,$\textnormal{d}_{xy}$ (14.03\%), while Se:\,$\textnormal{p}_{x}$ and Se:\,$\textnormal{p}_{z}$ contributed only modestly (6.69\% and 8.00\%, respectively); V2-derived contributions were absent at the CBM and VBM. In contrast, at the VBM (spin-down) the weight redistributed toward a mixed V–Se character: V1:\,$\textnormal{d}_{z^2}$ (39.27\%) remained significant, Se:\,$\textnormal{p}_{z}$ (27.47\%), Se:\,$\textnormal{p}_{y}$ (13.44\%) became prominent and V1:\,$\textnormal{d}_{x^2-y^2}$ (19.82\%) emerged, whereas V1:\,$\textnormal{d}_{zx}$ and V1:\,$\textnormal{d}_{xy}$ were absent. The orbitals of N atom showed negligible contributions in all cases.

\begin{table*}[t]
\caption{Orbital contributions for VSe$_2$/VN. Entries are percentages for the spin-up channel near $\mathrm{E_F}$ and for the spin-down channel at the CBM and VBM}
\label{tab:orbital-contrib-transposed}
\centering
\setlength{\tabcolsep}{4pt}
\scriptsize
\begin{tabularx}{\textwidth}{l *{9}{>{\centering\arraybackslash}X}}
\toprule
 & \textbf{$\mathrm{V1}:\,\textnormal{d}_{z^{2}}$}
 & \textbf{$\mathrm{V1}:\,\textnormal{d}_{zx}$}
 & \textbf{$\mathrm{V1}:\,\textnormal{d}_{xy}$}
 & \textbf{$\mathrm{V1}:\,\textnormal{d}_{x^{2}-y^{2}}$}
 & \textbf{$\mathrm{V2}:\,\textnormal{d}_{z^{2}}$}
 & \textbf{$\mathrm{V2}:\,\textnormal{d}_{xy}$}
 & \textbf{$\mathrm{Se}:\,\textnormal{p}_{x}$}
 & \textbf{$\mathrm{Se}:\,\textnormal{p}_{y}$}
 & \textbf{$\mathrm{Se}:\,\textnormal{p}_{z}$} \\
\midrule
\makecell[l]{\textbf{Spin-up channel near $\mathrm{E_F}$}}
 & 17.11\% & --- & --- & --- & 42.37\% & 10.01\% & --- & 5.07\% & 25.44\% \\ \\
\makecell[l]{\textbf{Spin-down channel at CBM}}
 & 51.70\% & 19.58\% & 14.03\% & --- & --- & --- & 6.69\% & --- & 8.00\% \\ \\
\makecell[l]{\textbf{Spin-down channel at VBM}}
 & 39.27\% & --- & --- & 19.82\% & --- & --- & --- & 13.44\% & 27.47\% \\ \\
\bottomrule
\end{tabularx}
\end{table*}


\begin{figure*}[!h]
\centering
\includegraphics[width=0.8\textwidth]{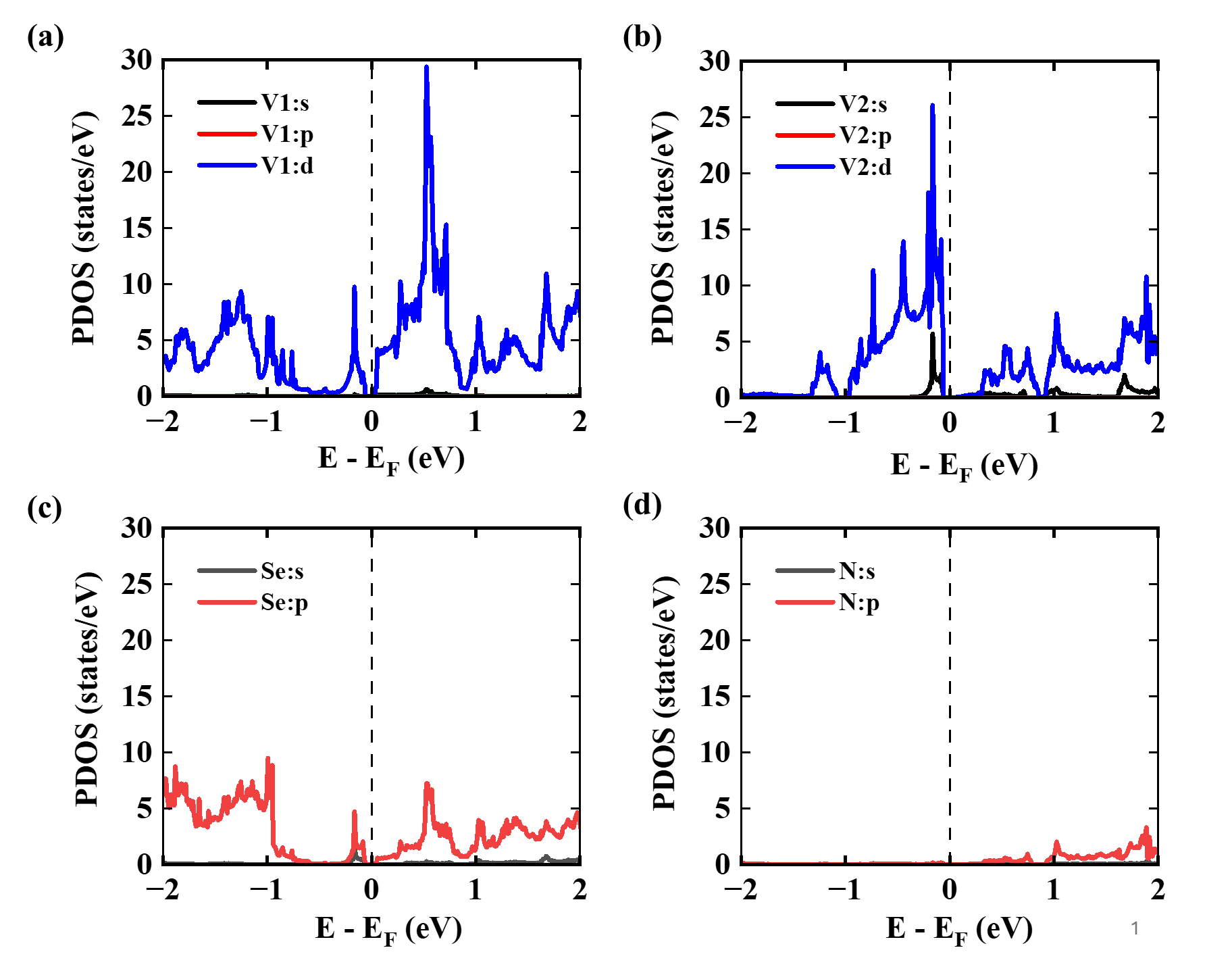}
  \caption{Orbital PDOS of VSe$_2$/VN ($2 \times 2 \times 1$ supercell). (a) Projection of  s, p, d orbitals of V1 (V atom of VSe$_2$). (b) Projection of s, p, d orbitals of V2 (V atom of VN).  (c) Projection of s, p orbitals of Se. (d) Projection of s, p orbitals of N.}
  \label{pdosorb}
\end{figure*}

\subsection{Magnetic Property of VSe$_2$/VN Heterostructure}

We found the total magnetization of VSe$_2$/VN to be  3.00~$\mu_{\mathrm{B}}$ per unit cell. The magnetocrystalline anisotropy energy (MAE) was calculated using the formula,
\begin{equation}
\hspace{2cm}\mathrm{MAE} = \mathrm{E}_{\parallel} - \mathrm{E}_{\perp}
\end{equation}
where $\mathrm{E}_{\parallel}$ is the total energy with the magnetization constrained in-plane (parallel to the layer) and $\mathrm{E}_{\perp}$ is the total energy with the magnetization constrained out-of-plane (perpendicular to the layer). With this sign convention, $\mathrm{MAE}>0$ indicates a perpendicular (out-of-plane) easy axis, while $\mathrm{MAE}<0$ indicates an in-plane easy axis.

For the unstrained case, the MAE
was \(+0.93\)~meV, indicating an out-of-plane easy axis. Under compressive (\(-4\%\)) and tensile (\(+4\%\)) strain, the MAE values were \(-1.279\) and \(+1.247\)~meV, respectively. Thus, compressive strain favors an in-plane easy axis, whereas both the unstrained and tensile cases favor an out-of-plane easy axis. The tensile case further strengthens the out-of-plane preference compared with 0\% strain.

We determined Curie temperature, $\mathrm{T}_{\mathrm{C}}$ using mean field estimation as mentioned in the study~\cite{Turek_2003}.
\begin{equation}
\hspace{1cm}\mathrm{T}_{\mathrm{C}}
= \frac{2\bigl(\mathrm{E}_{\mathrm{AFM}}-\mathrm{E}_{\mathrm{FM}}\bigr)}{3\,\mathit{k_B}}
\end{equation}

where $\mathit{k_B}$ is the Boltzmann constant, $\mathrm{E}_{\mathrm{AFM}}$ and $\mathrm{E}_{\mathrm{FM}}$ are the total energies of the antiferromagnetic and ferromagnetic states per magnetic atom, respectively. The Curie temperature ($\mathrm{T}_{\mathrm{C}}$) of VSe$_2$/VN was found to be 284.04~K, which is very close to room temperature (298 K). This indicates that the ferromagnetic order in VSe$_2$/VN should be thermally stable under ordinary laboratory conditions, making the system promising for room temperature spintronic applications. At compressive ($-4\%$) and tensile ($+4\%$) strain, the values of $\mathrm{T}_{\mathrm{C}}$ were found to be 47.5~K and 183.9~K, respectively, highlighting the strong strain sensitivity of the magnetic ordering and suggesting that near-ambient operation is most favorable close to the unstrained state.

Figs.~S3(a--e) illustrate the spin density, $\Delta \rho_{\mathrm{s}}(\mathbf{r})$, under different strain conditions. Negative spin density was marked in cyan and positive spin density was marked in yellow. In the compressive strain regime ($-2\%$ and $-1\%$), the negative spin density was weakened, whereas in the unstrained case and in the tensile regime ($+1\%$ and $+2\%$), the negative spin density strengthened and then saturated. At tensile strain, the structure was found to have constant  $\Delta \rho_{\mathrm{s}}$. As a result, the total magnetization is saturated at $3.00~\mu_{\mathrm{B}}$ per unit cell, as illustrated in Fig.~\ref{strainprof}(c).

\subsection{Effect of Biaxial Strain on Band Structure}
\begin{figure*}[!t]
\centering
\includegraphics[width=\textwidth]{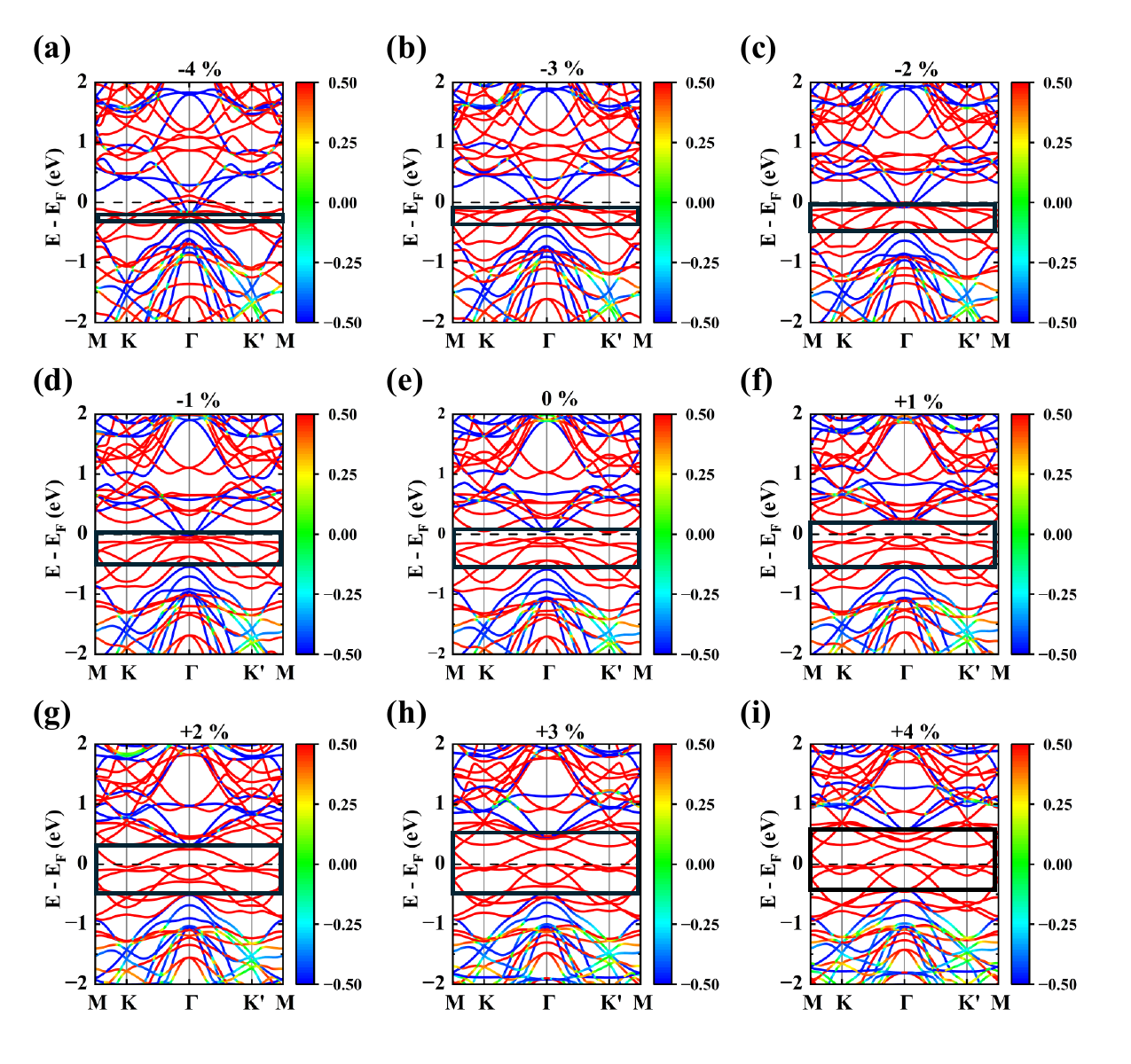}
\caption{Effect of biaxial strain on band structure: (a) $-4\%$, (b) $-3\%$, (c) $-2\%$, (d) $-1\%$, (e) $0\%$, (f) $+1\%$, (g) $+2\%$, (h) $+3\%$ and (i) $+4\%$. Panels (a)–(d) represent compressive strain, while (e)–(i) show tensile strain. The rectangular box shows the increasing gap of the spin-down channel.}
\label{BS}
\end{figure*}

Figs.~\ref{BS}(a–i) show the evolution of the band structure under biaxial strain. Under compressive strain and in the unstrained case (Figs.~\ref{BS}(a–e)), both spin channels retained a finite band gap, whereas under tensile strain (Figs.~\ref{BS}(f–i)), only the spin-down channel exhibited an increasing gap, while the spin-up gap collapsed. This behavior in VSe$_2$/VN enables enhanced spin selectivity near the Fermi level. A strain-controlled $|\mathrm{\Delta C_{KK'}}|$ was also observed in both spin channels. For the spin-down channel, $|\mathrm{\Delta C_{KK'}}|$ varied substantially with strain, whereas for the spin-up channel, $|\mathrm{\Delta C_{KK'}}|$ remained nearly constant as the strain was increased from compressive to tensile.

\begin{figure*}[h]
\centering
\includegraphics[width=\textwidth]{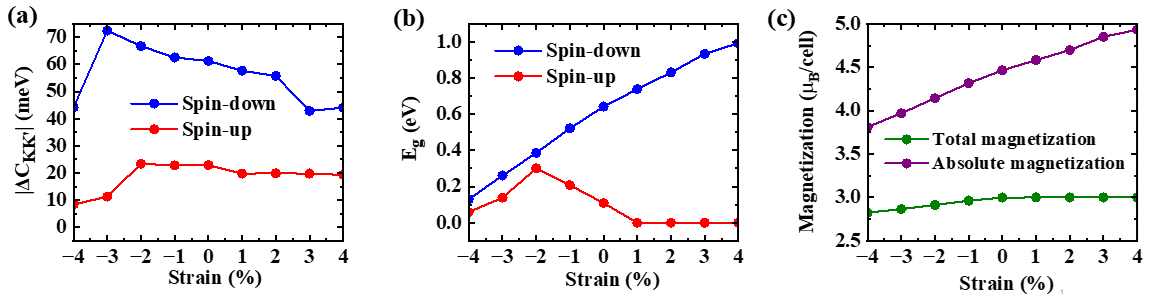}
  \caption{(a) Magnitude of conduction band valley splitting, $|\mathrm{\Delta C_{KK'}}|$, (b) bandgap, $\mathrm{E_g}$ of spin-down and spin-up channels, and (c) total and absolute magnetization  with respect to various strains (both compressive and tensile) of VSe$_2$/VN heterostructure.}
  \label{strainprof}
\end{figure*}

We plotted these observations in Figs.~\ref{strainprof}(a) and (b). In Fig.~\ref{strainprof}(a), $|\mathrm{\Delta C_{KK'}}|$ is shown with respect to strain for both spin channels. The spin-down $|\mathrm{\Delta C_{KK'}}|$ exhibited a pronounced, non-monotonic response: it increased from $44.2$~meV at $-4\%$ to a maximum of $72.5$~meV at $-3\%$, then gradually decreased through zero strain ($61.3$~meV) to a minimum of $42.9$~meV at $+3\%$, followed by a slight rebound to $44.1$~meV at $+4\%$. Overall, the spin-down splitting varied over a wide range ($\sim30$~meV, $\sim50\%$ of the zero-strain value), indicating strong tunability by biaxial strain. In contrast, the spin-up $|\mathrm{\Delta C_{KK'}}|$ remained comparatively small and weakly strain-dependent: it increased from $8.4$~meV at $-4\%$ to $23.4$~meV at $-2\%$ and then saturated near $\sim20$~meV for $-1\%\!\to\!+4\%$ (ending at $19.3$~meV). These trends show that strain chiefly modulates the spin-down valley splitting, while the spin-up channel is relatively insensitive.

Fig.~\ref{strainprof}(b) summarizes the spin-resolved band gaps (in eV) versus biaxial strain. The spin-down gap increased monotonically from $0.13$~eV at $-4\%$ to $0.99$~eV at $+4\%$, passing through $0.64$~eV at zero strain. In contrast, the spin-up gap showed a non-monotonic trend: it rose under compression, peaking at $0.35$~eV at $-2\%$, then decreased to $0.11$~eV at $0\%$ and collapsed to $0$ for tensile strains $\geq +1\%$. Thus, tensile strain drives a spin-selective regime in which the spin-down channel remains semiconducting with an enlarging gap, whereas the spin-up channel becomes gapless, consistent with a half-metallic behavior under tension.

Fig.~\ref{strainprof}(c) lists the strain dependence of the total and absolute magnetizations (in $\mu_{\mathrm{B}}$ per unit cell). The total magnetization increased steadily from $2.83~\mu_{\mathrm{B}}$ at $-4\%$ to $2.99~\mu_{\mathrm{B}}$ at $0\%$, and then saturated at $3.00~\mu_{\mathrm{B}}$ for tensile strains $\geq +1\%$. In contrast, the absolute magnetization rose almost monotonically across the entire range, from $3.81~\mu_{\mathrm{B}}$ at $-4\%$ to $4.93~\mu_{\mathrm{B}}$ at $+4\%$. This behavior is fully consistent with the previously discussed spin density of VSe$_2$/VN at various strains..


\subsection{Berry Curvature and Anomalous Hall Effect}

In condensed matter physics, Berry curvature acts as an effective
magnetic field in momentum space for Bloch electrons when TRS or SIS is broken~\cite{Berry_xiao, Berry_chang}. The Berry curvature in the conduction band is given by~\cite{schaibley2016valleytronics},
\begin{equation}
\hspace{1cm}
\mathrm{\Omega}_{\mathrm{c},\mathrm{s}}(\mathrm{k})
= -\,\tau_{z}\,
\frac{2\,\mathrm{a}^{2}\mathrm{t}^{2}\,\Delta}
{\bigl(4\,\mathrm{a}^{2}\mathrm{t}^{2}\,\mathrm{k}^{2}+\Delta^{2}\bigr)^{3/2}} 
\end{equation}
Here, \(\mathrm{k}\) is the crystal momentum measured from $\mathrm{K}$ or $\mathrm{K}'$, \(\Delta\) is the bandgap at \(\mathrm{k}\){, \(\mathrm{t}\) is the hopping integral, \(\mathrm{a}\) is the lattice constant, and \(\tau_{z}=\pm 1\) is the valley index.

In the presence of an in-plane electric field, E, this Berry curvature directly drives transverse anomalous velocities  according to
\begin{equation}
\hspace{1.5cm}
\mathrm{v}_{\mathrm{transverse}}
= \frac{\mathrm{e}}{\hbar}\,\mathrm{E}\times \mathrm{\Omega}(\mathrm{k}) 
\label{eq:velocity}
\end{equation}

Due to broken TRS or SIS, the Berry curvature will be opposite and unequal at $\mathrm{K}$ and $\mathrm{K}'$ valleys. As a result, $\mathrm{v}_{\mathrm{transverse}}$ will be opposite and asymmetric at these valleys, resulting in a generation of voltage known as anomalous Hall voltage.

The anomalous Hall conductivity or AHC was obtained by integrating the Berry curvature over all
occupied states in the Brillouin zone,
\begin{equation}
\hspace{1cm}
\mathrm{\sigma}_{xy}
= \frac{\mathrm{e}^{2}}{\hbar}\,
\int_{\mathrm{BZ}} \frac{d^{2}\mathrm{k}}{(2\pi)^{2}}\,
\mathrm{f}(\mathrm{k})\,\mathrm{\Omega}(\mathrm{k}) 
\label{eq:ahc}
\end{equation}
where $\mathrm{f}(\mathrm{k})$ is the Fermi–Dirac distribution function. VSe$_2$/VN heterostructure has both broken TRS and SIS. As illustrated in Fig.~\ref{fig:BC}(a), a non-zero net Berry curvature is observed at $\mathrm{K}$ and $\mathrm{K}'$. At the $\mathrm{K}$ valley, $-\mathrm{\Omega}_{\mathrm{z}}$ was $-9.05~\text{\AA}^{2}$, whereas at the $\mathrm{K}'$ valley it was $6.24~\text{\AA}^{2}$, confirming a finite net Berry curvature. Smaller features appeared around $\Gamma$, and between $\mathrm{K}$ and $\Gamma$, with several zero crossings indicated by the dashed baseline. A two-dimensional contour map of the Berry curvature over the hexagonal Brillouin zone is shown in Fig.~\ref{fig:BC}(b). The alternating blue and red hot spots at $\mathrm{K}$ and $\mathrm{K}'$ provide direct evidence for the presence of the anomalous Hall effect (AHE) in this heterostructure.

In Fig.~\ref{fig:BC}(c), the AHC denoted by $\mathrm{\sigma}_{\mathrm{xy}}$, is shown. For the spin-up channel, the maximum $\mathrm{\sigma}_{\mathrm{xy}}$ within the $\mathrm{K}$/$\mathrm{K}'$ energy window was $568.33$~S/cm, whereas for the spin-down channel it was $216.45$~S/cm. Because the spin-up bands were closer to $\mathrm{E_F}$, the AHC was dominated by the spin-up contribution. This large value of AHC underscores the potential for realizing robust AHE in VSe$_2$/VN.

\begin{figure*}[h]
    \centering
    \includegraphics[width=\textwidth]{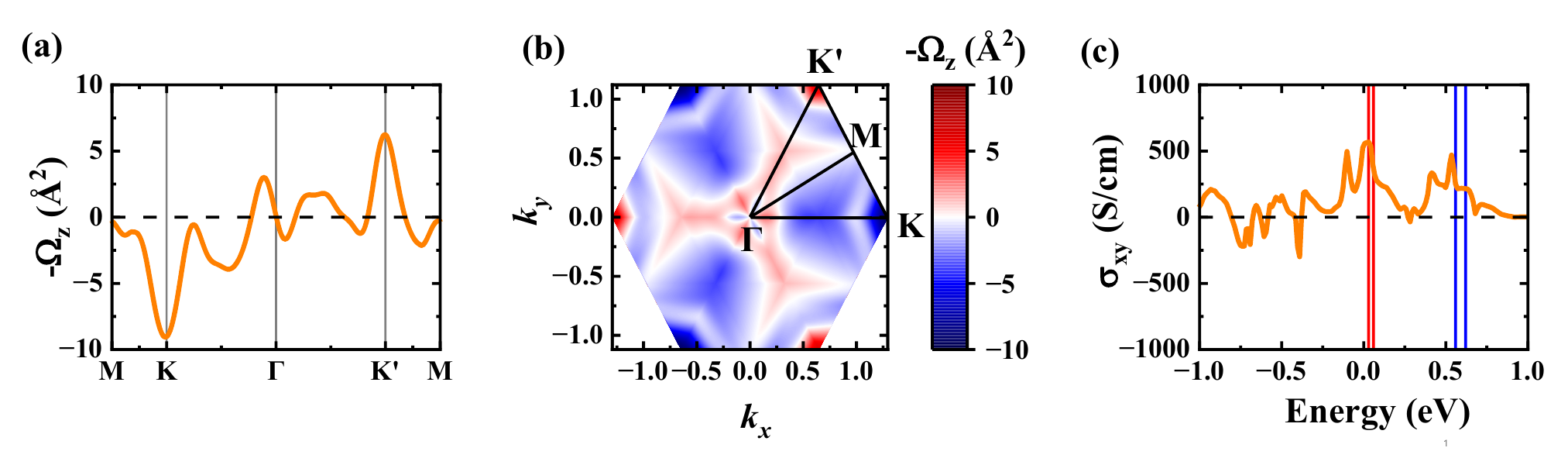}
    \caption{
    (a) Berry curvature, $-\mathrm{\Omega}_{\mathrm{z}}\,(\text{\AA}^2)$
    along the high-symmetry path $\mathrm{M} \to \mathrm{K} \to \mathrm{\Gamma} \to \mathrm{K'} \to \mathrm{M}$, showing valley-contrasting behavior at $\mathrm{K}$ and $\mathrm{K'}$ points.
    (b) A 2D contour plot of Berry curvature, $-\mathrm{\Omega}_{\mathrm{z}}\,(\text{\AA}^2)$ showing opposite valley index at alternating $\mathrm{K}$ and $\mathrm{K'}$ points over the whole hexagonal Brillouin zone (high-symmetry path is marked in the figure).
    (c) Anomalous Hall conductivity, $\mathrm{\sigma}_{\mathrm{xy}}$ of VSe$_2$/VN (red and blue lines indicate the $\mathrm{K}$ and $\mathrm{K'}$ valleys' energy range of the spin-up and spin-down band structures' CBM, respectively).
    }
    \label{fig:BC}
\end{figure*}

\subsection{Proposed Tunable Spin-Filter and Valleytronic Device}

To evaluate the spin-filtering efficiency of VSe$_2$/VN, we followed the low-bias Wentzel–Kramers–Brillouin (WKB) description of spin-dependent tunneling as mentioned in previous literatures~\cite{simmons1963generalized,TedrowMeservey}. The spin-resolved current densities scale exponentially with the decay constants inside the barrier,
\begin{equation}
\mathrm{J}_{\uparrow,\downarrow}\;\propto\;
\exp\!\big[-2\,\kappa_{\uparrow,\downarrow}\,d\big],
\qquad
\kappa_{\uparrow,\downarrow}
=\frac{\sqrt{2\,m^{\ast}\,\Phi_{\uparrow,\downarrow}}}{\hbar} 
\label{eq:wkb}
\end{equation}
where the spin-dependent barrier heights are written as symmetric deviations about their mean,
\begin{equation}
\Phi_{\uparrow,\downarrow}
=\bar{\Phi}\mp\frac{\Delta\Phi}{2},
\qquad
\Delta\Phi=\Phi_{\downarrow}-\Phi_{\uparrow} 
\label{eq:phi_split}
\end{equation}
The spin-filtering efficiency is then defined as
\begin{equation}
\hspace{2cm}
\mathrm{P}
=\frac{\mathrm{J}_{\uparrow}-\mathrm{J}_{\downarrow}}
{\mathrm{J}_{\uparrow}+\mathrm{J}_{\downarrow}} 
\label{eq:P_def}
\end{equation}

Here, $\mathrm{J}_{\uparrow,\downarrow}$ denotes spin-up and spin-down tunneling current densities, $d$ is the barrier thickness,
$m^{\ast}$ is the electron effective mass and 
$\Phi_{\uparrow,\downarrow}$ is the spin-dependent barrier heights.

Fig.~\ref{final}(a) demonstrates spin-filter tunability of VSe$_2$/VN under biaxial tensile strain. Biaxial strain can be applied using wafer-bending method, as described in the literature~\cite{kumar2024strain}. At $0\%$ and $+4\%$ strain, the gaps were $0.64$ and $0.99$~eV, respectively; this gap served as the tunneling barrier for the spin-down channel. Using the spin-filtering efficiency expression, we obtained $\mathrm{P}=75.4\%$ for the unstrained case and $\mathrm{P}=82.5\%$ at $+4\%$ tensile strain, indicating a clear improvement. We focused on $+4\%$ because it yielded a higher DOS near $\mathrm{E_F}$ than in the unstrained case (Figs.~S5(a) and (b)), with an $\sim8.4$ times enhancement of spin-up carriers (Section~S6 of Supplementary Material). Thus, tensile strain simultaneously increases both efficiency and spin-up carrier concentration. The trade-off is a reduction of the Curie temperature from $284.04$~K to $183.9$~K, implying that cryogenic operation would be required for robust ferromagnetism at $+4\%$ strain.

A valleytronic transistor based on VSe$_2$/VN can be realized with source, drain, a back gate, and lateral probes A and B.  An analogous valleytronic transistor was demonstrated for MoS$_2$ on SiO$_2$/Si in the study~\cite{li2020room}. Figs.~\ref{final}(b) and \ref{final}(c) show the top and side views of the proposed device. When a drain–source bias $\mathrm{V_{DS}}$ will be applied, electrons will flow along the channel; the opposite Berry curvatures at $\mathrm{K}$ and $\mathrm{K}'$ will deflect carriers to opposite edges, producing an anomalous Hall voltage $\mathrm{V_H}$ between probes A and B. The back-gate voltage $\mathrm{V_G}$ will tune the Fermi level $\mathrm{E_F}$ and thereby modulate $\mathrm{V_H}$.


\begin{figure*}[h]
\centering
 \includegraphics[width=\textwidth]{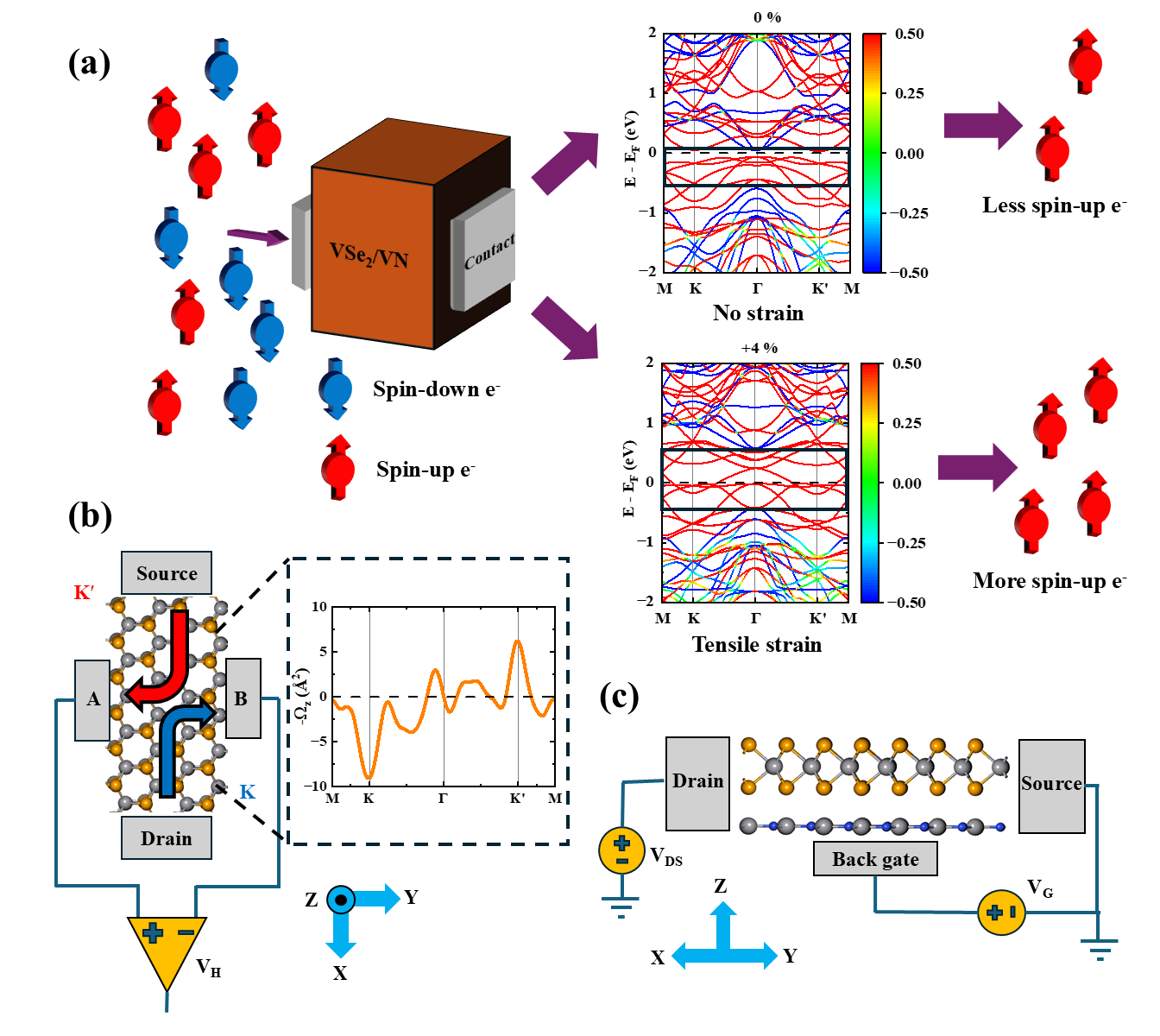}
  \caption{(a) Tunability of spin-filter. Tensile strain favors high spin-filtering efficiency along with high spin-up electron concentration. (b) Top and (c) side view of the proposed valleytronic device with drain, source, back gate and probes A and B. Anomalous Hall voltage, $\mathrm{V_H}$ will appear between A and B.}
  \label{final}
\end{figure*}

\section{Comparative Analysis}

In Table~\ref{tab:valley-AHC}, VSe$_2$/VN shows a balanced and superior performance across all four metrics: Curie temperature $\mathrm{T}_{\mathrm{C}}$, spin-filtering efficiency $\mathrm{P}$, conduction band valley splitting $\mathrm{\Delta C_{KK'}}$, and maximum anomalous Hall conductivity $|\mathrm{\sigma}_{xy}|$ evaluated between $\mathrm{K}$ and $\mathrm{K}'$ energy. Specifically, VSe$_2$/VN attained $\mathrm{T}_{\mathrm{C}}=284.04~\mathrm{K}$ with $\mathrm{P}=75.4\%$, exceeding W$_{0.25}$V$_{0.75}$Se$_2$ $(59.57~\mathrm{K},\,42\%)$~\cite{Shihab2022ACSOmega}, EuO $(69.3~\mathrm{K},\,29\%)$~\cite{inomata2008highly} and EuS $(16.6~\mathrm{K},\,86\%)$~\cite{inomata2008highly} by providing near-ambient $\mathrm{T}_{\mathrm{C}}$ while maintaining high $\mathrm{P}$. For valley properties, VSe$_2$/VN exhibited $\mathrm{\Delta C_{KK'}}=22.9~\mathrm{meV}$, comparable to V-doped AgBiP$_2$S$_6$ $(26.8~\mathrm{meV})$~\cite{ZhangZhou} and NbSe$_2$ $(18.6~\mathrm{meV})$~\cite{Samiul2025APSUSC} and below GdI$_2$ $(149~\mathrm{meV})$~\cite{cheng2021two}. The AHC of VSe$_2$/VN outperformed the other systems. It reached $|\mathrm{\sigma}_{xy}|=568.33~\mathrm{S/cm}$, larger in magnitude than NbSe$_2$ $(\sim\!59~\mathrm{S/cm})$, V-doped AgBiP$_2$S$_6$ $(\sim\!120~\mathrm{S/cm})$ and GdI$_2$ $(\sim\!13~\mathrm{S/cm})$. Entries for W$_{0.25}$V$_{0.75}$Se$_2$, EuO, and EuS remain absent for $\mathrm{\Delta C_{KK'}}$ and $\mathrm{\sigma}_{xy}$, indicating no reported valley-resolved Hall data. Taken together, these comparisons show that only this study (VSe$_2$/VN) simultaneously demonstrated robust spin filtering and clear valley-related phenomena; namely a finite $\mathrm{\Delta C_{KK'}}$ in the conduction band together with a large $\mathrm{\sigma}_{xy}$ between $\mathrm{K}$ and $\mathrm{K}'$ which constitutes the central novelty of our work.

\newcolumntype{Y}{>{\centering\arraybackslash}X}
\newcolumntype{Z}{>{\centering\arraybackslash}p{0.28\textwidth}} 

\begin{table*}[t]
\caption{Comparative study of 2D systems exhibiting spin-filtering, valley properties and anomalous Hall effect}
\label{tab:valley-AHC}
\centering
\setlength{\tabcolsep}{4pt}
\renewcommand{\arraystretch}{1.15} 
\small
\begin{tabularx}{\textwidth}{l Y Y Y Z l}
\toprule
\textbf{Structure} &
\textbf{Curie Temperature, $\mathrm{T}_{\mathrm{C}}$ (K)} &
\textbf{Spin-filtering efficiency, $\mathrm{P}$ (\%)} &
\textbf{Conduction band valley splitting, $\mathrm{\Delta C_{KK'}}$ (meV)} &
\textbf{Maximum anomalous Hall conductivity, $|\mathrm{\sigma}_{xy}|$ (S/cm) between $\mathrm{K}$ and $\mathrm{K}'$ energy} &
\textbf{Reference} \\
\midrule
W$_{0.25}$V$_{0.75}$Se$_2$ & 59.57  & 42   & --    & --             & ~~~\cite{Shihab2022ACSOmega} \\
EuO                         & 69.3   & 29   & --    & --             & ~~~~\cite{inomata2008highly} \\
EuS                         & 16.6   & 86   & --    & --             & ~~~~\cite{inomata2008highly} \\
VSe$_2$/VN                  & 284.04 & 75.4 & 22.9 & 568.33        & This study \\
NbSe$_2$                    & 176.25 & --    & 18.6 & \hspace{-0.3cm}$\sim 59$ & ~~~\cite{Samiul2025APSUSC} \\
V-doped AgBiP$_2$S$_6$      & --      & --    & 26.8 & \hspace{-0.3cm}$\sim 120$   & ~~~\cite{ZhangZhou} \\
GdI$_2$                     & --      & --    & 149  & \hspace{-0.3cm}$\sim 13$     & ~~~\cite{cheng2021two} \\
\bottomrule
\end{tabularx}
\end{table*}

\section*{Conclusion}

In this paper, we explored the VSe$_2$/VN van der Waals heterostructure and established that the AA-stacked configuration (stacking~2) was the most stable, exhibiting a formation energy of $-0.95$~eV per unit cell and an in-plane lattice parameter of $3.328$~\AA. We further found that charge transferred from the VN layer to the VSe$_2$ layer upon forming the heterostructure, consistent with the work functions of VSe$_2$ (5.47~eV), VN (3.81~eV), and the combined system (4.53~eV), which aligned at the interface. The optimized structure yielded Se--V--Se and V--N--V/N--V--N bond angles of $79.29^\circ$ and $120^\circ$, respectively, confirming minimal distortion of the constituent layers. Electrically and magnetically, the heterostructure displayed key ferrovalley features with a sizable and strain-tunable conduction band valley splitting in both spin channels ($\mathrm{\Delta C_{KK'}}=22.9~\mathrm{meV}$ for spin-up and  $\mathrm{\Delta C_{KK'}}=61.3~\mathrm{meV}$ for spin-down) including total magnetization of 3.00~$\mu_{\mathrm{B}}$ per unit cell, out-of-plane easy axis at unstrained and tensile strain cases, and strong Berry-curvature–driven transport, culminating in a maximum anomalous Hall conductivity of $568.33$~(S/cm). Importantly, the heterostructure had a Curie temperature of $284.04$~K and a spin-filtering efficiency of $75.4\%$ in the unstrained case and tensile strain-tunable up to $82.5\%$ with a sacrifice in Curie temperature ($183.9$~K), demonstrating concurrent prospects for valleytronic and spintronic device concepts in a single material platform. These results revealed that VSe$_2$/VN combined robust thermodynamic stability with interfacial charge redistribution and spin–valley functionality, positioning it as a promising candidate for next-generation spin-filtering and valleytronic applications.

\printcredits
\section*{Declaration of Competing Interest}
The authors declare that they have no known competing financial interests or personal relationships that could have appeared to influence the work reported in this paper.

\section*{Data Availability Statement}
{The data supporting the findings presented in this paper are not currently available to the public, but they may be obtained from the authors upon reasonable request.} 

\section*{Acknowledgements}
V.C. and A.Z. thank the Bangladesh University of Engineering and Technology (BUET) and the Bangladesh Research and Education Network (BdREN) for providing computational facilities and technical support. A.Z. and V.C. further thank the Research and Innovation Centre for Science and Engineering (RISE), BUET, for financial support through the Internal Research Grant (Project ID: 2023-02-022).

\section*{Supplementary Material}
See Supplementary Material at https://....

\balance

\bibliographystyle{unsrt}
\bibliography{cas-refs}





\end{document}